\documentclass[reprint,amsmath,amssymb,aps,prb,superscriptaddress]{revtex4-2}

\usepackage[T2A]{fontenc}
\usepackage[utf8]{inputenc}
\usepackage{soul}
\usepackage{xcolor}
\usepackage{graphicx}
\usepackage{dcolumn}
\usepackage{bm}

\usepackage{placeins}

\begin{document}

\preprint{APS/123-QED}

\title{Degeneracy and trajectory control of spin eigenmodes excited by fs-optical pulses in a nearly compensated ferrimagnet}

\author{G.\,Yu.~Levkin}
\affiliation{Lomonosov Moscow State University, Faculty of Physics, 121205 Moscow, Russia}
\affiliation{Russian Quantum Center, 121353 Moscow, Russia}
\author{D.\,M.~Krichevsky}
\affiliation{Lomonosov Moscow State University, Faculty of Physics, 121205 Moscow, Russia}
\affiliation{Russian Quantum Center, 121353 Moscow, Russia}
\author{N.\,A.~Gusev}
\affiliation{Lomonosov Moscow State University, Faculty of Physics, 121205 Moscow, Russia}
\affiliation{Russian Quantum Center, 121353 Moscow, Russia}
\author{A.\,K.~Zvezdin}
\affiliation{Lomonosov Moscow State University, Faculty of Physics, 121205 Moscow, Russia}
\affiliation{Russian Quantum Center, 121353 Moscow, Russia}
\affiliation{Prokhorov General Physics Institute of the Russian Academy of Sciences, 119991 Moscow, Russia}
\author{S.\,N.~Polulyakh}
\affiliation{Physics and Technology Institute, Vernadsky Crimean Federal University, 295007 Simferopol, Crimea}
\author{V.\,I.~Belotelov}
\affiliation{Lomonosov Moscow State University, Faculty of Physics, 121205 Moscow, Russia}
\affiliation{Russian Quantum Center, 121353 Moscow, Russia}
\author{D.\,O.~Ignatyeva}
\affiliation{Lomonosov Moscow State University, Faculty of Physics, 121205 Moscow, Russia}
\affiliation{Russian Quantum Center, 121353 Moscow, Russia}

\date{\today}

\begin{abstract}
We investigate optically excited spin dynamics in a uniaxial ferrimagnet near the magnetization compensation point under a magnetic field applied along the magnetic anisotropy axis. Experiment and numerical modeling reveal an unusual regime where the frequencies of two spin eigenmodes approach each other and become highly field sensitive. The modes, corresponding to opposite rotations of the Néel vector, simultaneously reverse their handedness at a critical field where their frequencies become degenerate. At this point, the two-frequency precessional dynamics collapses into a linear oscillations directed along the inverse-Faraday-effect excitation induced by a single pump pulse. We further show that a double-pulse excitation scheme enables control of the spin trajectory. These results uncover an unconventional dynamical regime in ferrimagnets and establish new opportunities for manipulating spin motion in magnonic systems and devices.
\end{abstract}

\maketitle

\section{\label{sec:intro}Introduction}

Ultrafast magnetic phenomena driven by femtosecond light pulses are emerging topics of modern condensed matter physics~\cite{kirilyuk2010ultrafast, kalashnikova2015ultrafast, stupakiewicz2017ultrafast, kimel2019writing, barman20212021}. In transparent magnetic dielectrics, circularly polarized pulses can act as an ultrashort nonthermal stimulus through the inverse Faraday effect~\cite{pershan1966theoretical}, enabling coherent excitation and helicity-dependent control of spin motion without relying on strong optical absorption~\cite{kimel2005ultrafast, zhu2022inverse, gribova2024inverse}. Thus, light spatial profile tailoring an additional control of the excited spin dynamics parameters~\cite{satoh2012directional, okyay2020resonant, krichevsky2021selective, kolosvetov2022concept, hiraoka2024sublattice, krichevsky2024inverse, konkov2026selective}. Moreover, launching of spin dynamics via the pulse sequences instead of a single pulse provides an additional mean of the dynamics control~\cite{khramova2022accumulation, kolodny2019resonant}.

The spin dynamics of ferrimagnets differs qualitatively from that of single-sublattice ferromagnets. Ferrimagnets are magnetic materials composed of two or more magnetic sublattices. Owing to this sublattice structure, their net magnetization may vanish at the magnetization compensation temperature $T_M$, where the sublattice magnetizations compensate each other. The same two-sublattice nature also gives rise to two spin-precession modes~\cite{kaplan1953exchange, wangsness1953sublattice, kittel1959theory}. Far from $T_M$, these modes are described as the ferromagnetic resonance mode and the exchange mode. Ferromagnetic mode typically has the frequency of several GHz and linear dependence on the external magnetic field~\cite{kittel1948theory}. The frequency of the exchange mode is determined by the exchange interaction, so it usually has the values of about hundreds of GHz, and the impact of the external magnetic field on its frequency is quite weak~\cite{kaplan1953exchange, geschwind1959exchange}.

In the case the temperature is close to $T_M$ the frequencies, damping and motion of the eigenmodes become sensitive to the balance between exchange interaction, anisotropy, and value and orientation of external field~\cite{stanciu2006ultrafast, binder2006magnetization, ivanov2019ultrafast, davydova2020ultrafast, deb2022magnetic, xu2026inversion}. It was shown that in the vicinity of $T_M$ the exchange mode may acquire strong field sensitivity, whereas the ferromagnetic mode may lose it and soften near the phase transition between the collinear and non-collinear states~\cite{deb2016temperature, krichevsky2023unconventional, ignatyeva2025high, mikuni2025magnetic}.

In an in-plane field, the field competes with the perpendicular anisotropy and can drive the film between collinear and non-collinear states~\cite{krichevsky2023unconventional, wu2024magnon}. In the out-of-plane geometry considered here, the field is applied along the anisotropy axis, thus within the studied field range this geometry keeps the equilibrium state collinear. Geschwind and Walker studied gadolinium iron garnet near the magnetization compensation temperature and observed the ferromagnetic and exchange resonance branches, showing that their field dependences above and below compensation follow the two-sublattice resonance theory~\cite{geschwind1959exchange}. However, the motion of the eigenmodes were not addressed. Mikuni \textit{et al.} derived analytical expressions for resonance frequencies of two-sublattice ferrimagnets over the whole temperature range for both in-plane and out-of-plane orientations~\cite{mikuni2025magnetic}. Despite the precise analytical formulas obtained, the behavioral features of modes in the vicinity of $T_M$ and the influence of the magnetic field on them were not analyzed in detail. Thus, spin dynamics near compensation in the out-of-plane geometry remains only partially explored.

In the current work we study laser-induced spin dynamics in a nearly compensated uniaxial iron-garnet film with the external magnetic field applied normally to the film plane and along the anisotropy axis. Using a quasi-antiferromagnetic description of the two-sublattice system, we show that near $T_M$ the eigenmodes are clockwise and counterclockwise circular precessions of the Neel vector. Their frequencies depend linearly on the magnetic field with opposite slopes, so that at a certain field and temperature they become equal. At this point the usual two-mode precession degenerates into an oscillation along the direction set by the plane of incidence of the exciting pulse. We further show that a second femtosecond pulse provides trajectory control of the Neel-vector motion. Our results demonstrate control not only of mode frequencies but also of the trajectory in a compensated ferrimagnetic.

\section{Spin eigenmodes}\label{sec:model}

Experimental and theoretical studies of spin dynamics in a uniaxial ferrimagnet were performed exemplary on a ferrimagnetic iron-garnet film of composition $\mathrm{(Bi_{0.77}Y_{0.92}Lu_{1.31})(Fe_{3.45}Ga_{1.55})O_{12}}$ grown epitaxially on the (111) gadolinium gallium substrate is chosen. It has relatively large uniaxial magnetic anisotropy due its growth process. Magnetization of the tetrahedral magnetic sublattice of the film, $\mathbf{M_1}$, is diminished by dilution of $\mathrm{Fe}^{3+}$ ions with $\mathrm{Ga}^{3+}$ ions. This makes the value of $\mathbf{M_1}$ close to magnetization of the octahedral sublattice, $\mathbf{M_2}$. Due to the different temperature dependencies of $|\mathbf{M_1}|$ and $|\mathbf{M_2}|$ there is a temperature $T_M=333$~K at which the net magnetization becomes zero. 

\subsection{Equations of motion}

Theoretical and numerical studies of spin dynamics of a ferrimagnet in the vicinity of a magnetization compensation point were conducted using quasi-antiferromagnetic approach~\cite{davydova2020ultrafast, krichevsky2023unconventional}, in which magnetization vectors of the two magnetic sublattices $\mathbf{M_1}$ and $\mathbf{M_2}$ are considered nearly opposite to each other. The Lagrangian for the Neel vector $\mathbf{L}=\mathbf{M_1}-\mathbf{M_2}$ of such a ferrimagnet~\cite{krichevsky2023unconventional} placed in an external magnetic field directed along the anisotropy axis can $\mathbf{H}$ be written as:
\begin{equation}
\begin{split}
\mathcal{L}_\mathrm{eff} =& \tfrac{\chi_\perp}{2}\left[
  \left(\tfrac{\dot\varphi}{\gamma}\sin\theta + H\cos\theta\cos\varphi\right)^2
 +\left(\tfrac{\dot\theta}{\gamma} + H\sin\varphi\right)^2
\right]\\&
 - \tfrac{\dot\varphi}{\gamma} m\cos\theta
 + mH\sin\theta\cos\varphi
 + K\sin^2\theta\cos^2\varphi ,
\end{split}
\label{Eq: Lagrangian}
\end{equation}
where $\theta$ and $\varphi$ are the angles in spherical coordinate system that determine $\mathbf{L}$ orientation, $m=\left| \mathbf{M_1} \right|-\left| \mathbf{M_2} \right|$, $\gamma$ is a gyromagnetic constant, $K$ is a magnetic uniaxial anisotropy constant,  $\chi_\perp=\frac{\left(\left| \mathbf{M_1} \right|+\left| \mathbf{M_2} \right|\right)^2}{4\Lambda\left| \mathbf{M_1} \right|\left| \mathbf{M_2} \right|}$ is a perpendicular magnetic susceptibility, $\Lambda$ is a Weiss constant describing the exchange interaction of the two sublattices.

\begin{figure}[htb]
  (a)~~~~~~~~~~~~~~~~~~~~~~~~~~~~~~~~~~~~~~~~~~~~~~~~~~~~~~~~~~~~~~~~~~~~~~~~~~~~~~~~\\
  \includegraphics[width=\linewidth]{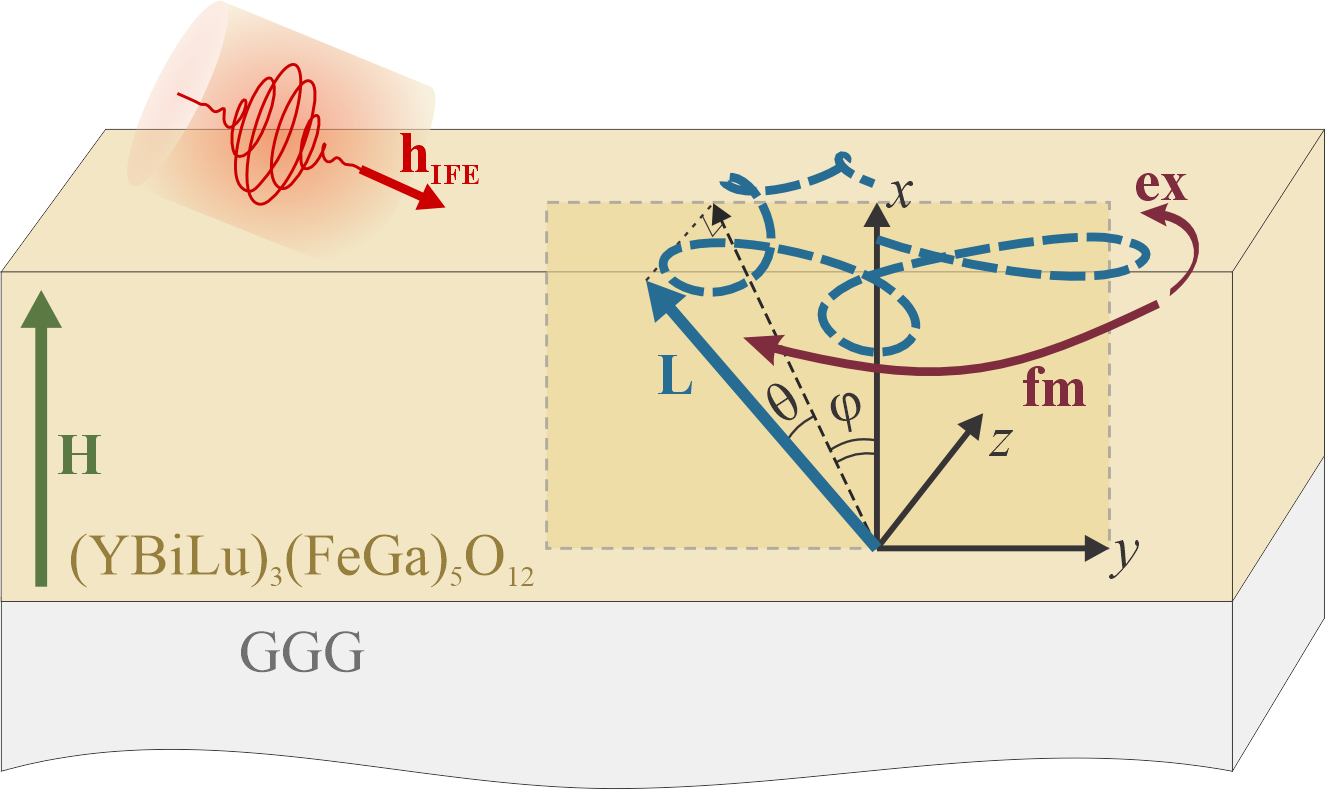}\\
  (b)~~~~~~~~~~~~~~~~~~~~~~~~~~~~~~~~~~~~~~~~~~~~~~~~~~~~~~~~~~~~~~~~~~~~~~~~~~~~~~~~\\
  \includegraphics[width=\linewidth]{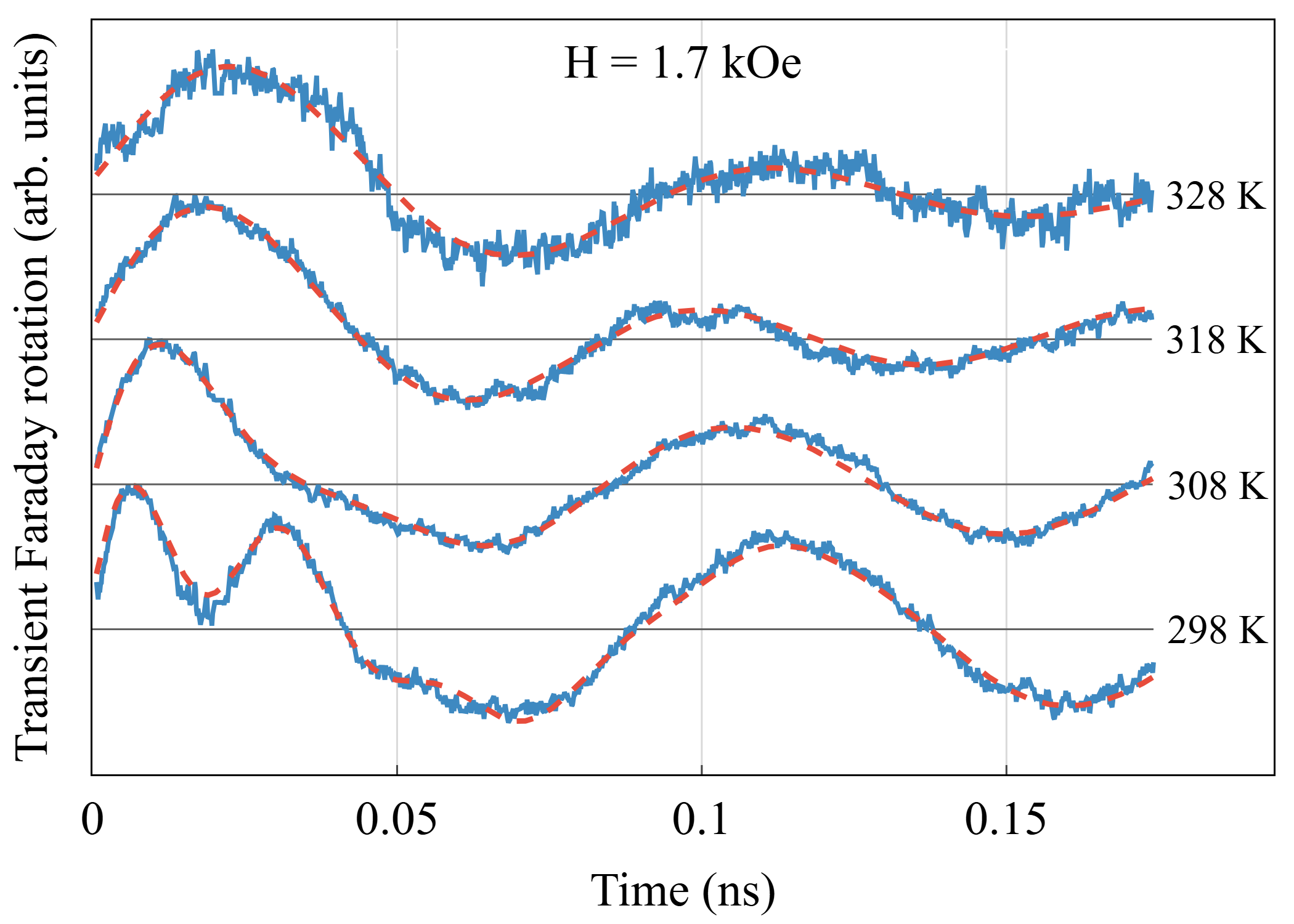}

  \caption{Spin precession in normally applied magnetic field. (a) Schematic representation of the configuration. (b) Normalized TFR signals and their approximations for the case $H=1.7$~kOe.}
  \label{Fig: config}
\end{figure}

\begin{figure*}[htb]
(a)~~~~~~~~~~~~~~~~~~~~~~~~~~~~~~~~~~~~~~~~~~~~~~~~~~~~~~~~~~~~~~~~~~~~~~~~~~~~~~~~~~~~~~(b)~~~~~~~~~~~~~~~~~~~~~~~~~~~~~~~~~~~~~~~~~~~~~~~~~~~~~~~~~~~~~~~~~~~~~~~~~~~~~~~~~~\\
  \includegraphics[width=0.49\linewidth]{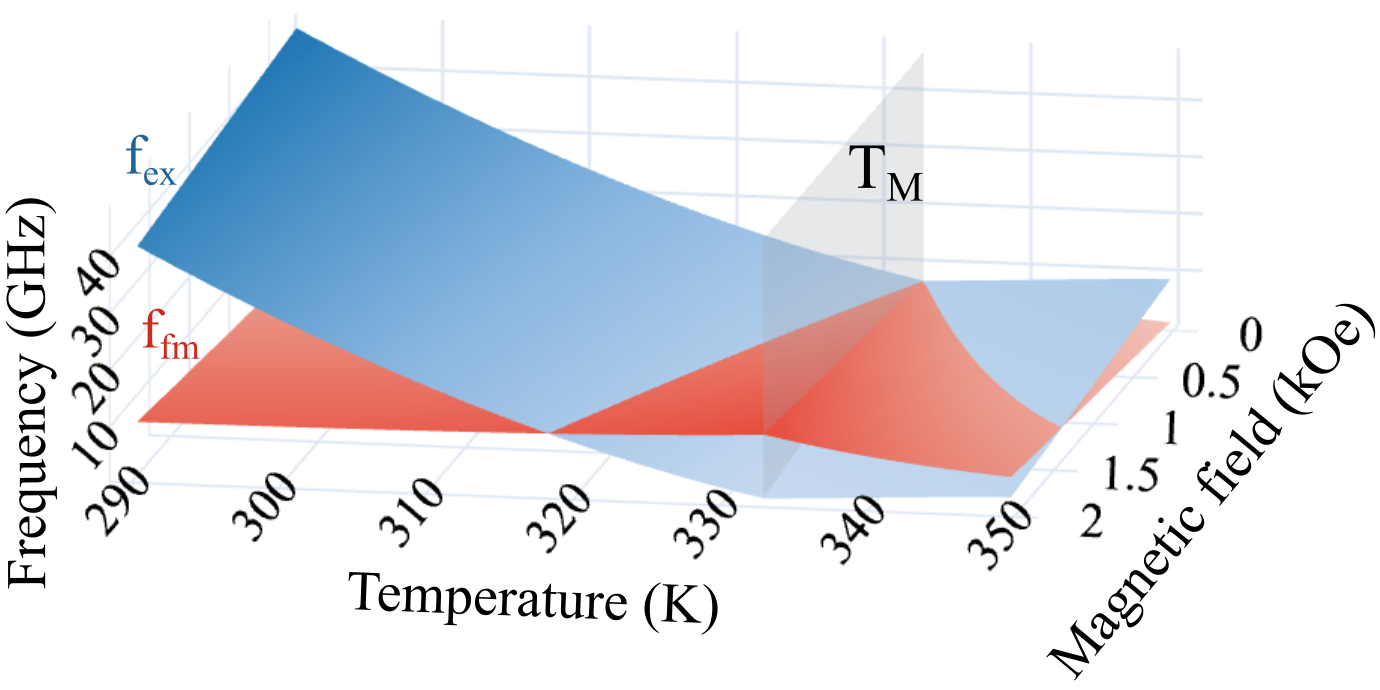}\hfill
  \includegraphics[width=0.49\linewidth]{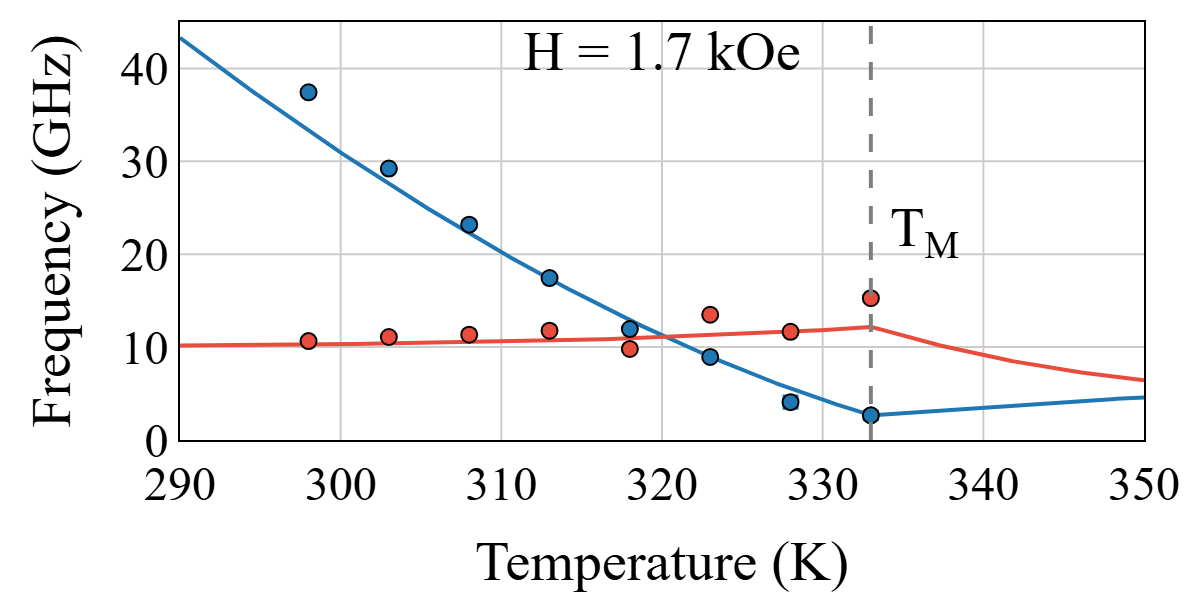}\\
(c)~~~~~~~~~~~~~~~~~~~~~~~~~~~~~~~~~~~~~~~~~~~~~~~~~~~~~~~~~~~~~~~~~~~~~~~~~~~~~~~~~~~~~~(d)~~~~~~~~~~~~~~~~~~~~~~~~~~~~~~~~~~~~~~~~~~~~~~~~~~~~~~~~~~~~~~~~~~~~~~~~~~~~~~~~~~\\

  \includegraphics[width=0.49\linewidth]{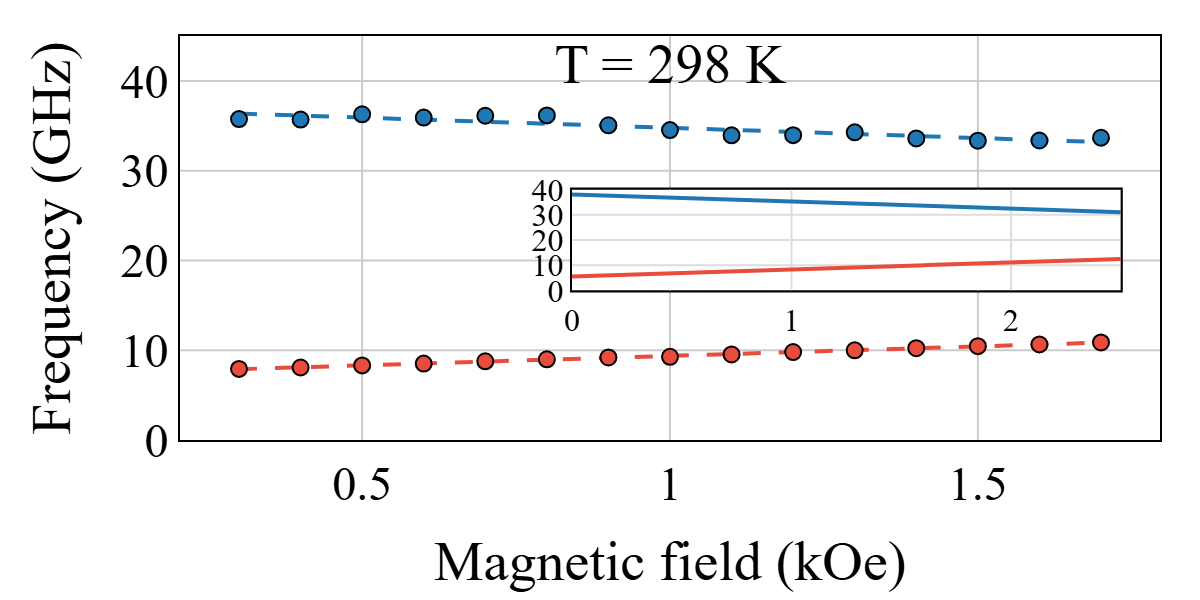}\hfill
  \includegraphics[width=0.49\linewidth]{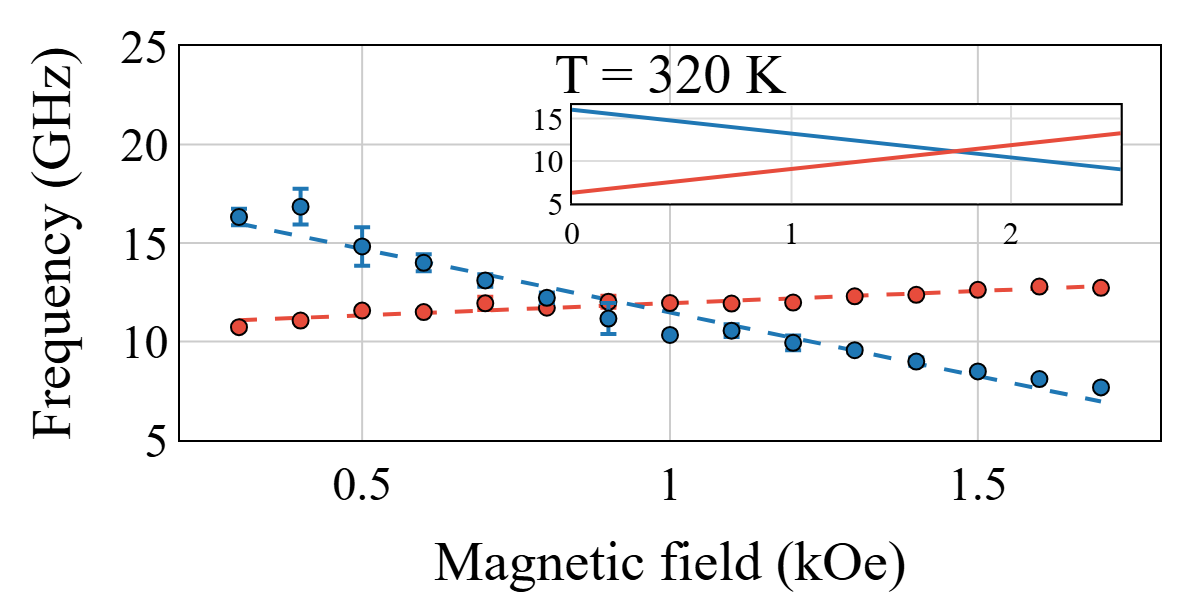}

  \caption{(a) Spin precession frequencies of exchange $f_\mathrm{ex}$ (blue color) and ferromagnetic $f_\mathrm{fm}$ (red color) modes in $(H,T)$ coordinates. (b-d) Experimental measurements of spin precession. (b) Temperature dependence of exchange and ferromagnetic mode frequencies for $H=1.7$~kOe. (c,d) Magnetic field dependence of exchange and ferromagnetic frequencies for (c) $T=298$~K  and (d) $T=320$~K. Dots represent experimental results, solid lines show theoretical curves, dashed lines are linear fits of experimental dots.}
  \label{Fig: freqs}
\end{figure*}

Linearized motion equations can be obtained from Eq.~\eqref{Eq: Lagrangian} as: 
\begin{equation}
    \begin{aligned}
        \ddot{\theta} + 2\Gamma\cdot\dot{\theta} + \Omega^2\cdot\theta \pm 2\delta\cdot\dot{\varphi} &= 0,\\
        \ddot{\varphi} + 2\Gamma\cdot\dot{\varphi} + \Omega^2\cdot\varphi \mp 2\delta\cdot\dot{\theta} &= 0,
    \end{aligned}
    \label{Eq: theta_and_phi}
\end{equation}
where $\Gamma=\frac{\alpha M\gamma}{2\chi_\perp}$ is a damping constant, $\alpha$ is a Gilbert damping constant, $M=\left| \mathbf{M_1} \right|+\left| \mathbf{M_2} \right|$, $\Omega^2=\omega_0^2-\delta^2$, $\omega_0^2=\omega_a^2+\tfrac{1}{4}\omega_{KK}^2$, $\delta = \omega_H-\frac{1}{2}\omega_{KK}$, $\omega_H=\gamma H$, $\omega_{KK}=\gamma\frac{\left|m\right|}{\chi_\perp}$, $\omega_a=\gamma\sqrt{2K/\chi_\perp}$. Signs $\pm$ and $\mp$ describe the cases of $m\geq0$ (upper sign) and $m<0$ (bottom sign). Such a difference in equations is caused by the different $\mathbf{L}$ orientation with respect to $\mathbf{H}$ below and above magnetization compensation point, which is caused by parallel $\mathbf{L}\uparrow\uparrow\mathbf{M}$ and antiparallel $\mathbf{L}\uparrow\downarrow\mathbf{M}$ alignment below and above $T_\mathrm{M}$.

Equations~\eqref{Eq: theta_and_phi} can be written in terms of complex coordinates describing the circular precession with opposite helicity of ferromagnetic $\Psi_\mathrm{fm} =\theta\mp i\varphi$ and exchange $\Psi_\mathrm{ex} =\theta\pm i\varphi$ modes, where the sign corresponds to the sign of $m$ as denoted above: 

\begin{equation}
    \begin{aligned}
        \ddot{\Psi}_\mathrm{fm} + 2\left(\Gamma+i\delta\right)\cdot\dot{\Psi}_\mathrm{fm} + \Omega^2\cdot\Psi_\mathrm{fm} &= 0,\\
        \ddot{\Psi}_\mathrm{ex} + 2\left(\Gamma-i\delta\right)\cdot\dot{\Psi}_\mathrm{ex} + \Omega^2\cdot\Psi_\mathrm{ex} &= 0.
    \end{aligned}
    \label{Eq: modes}
\end{equation}

It is important that Eqs.~\eqref{Eq: modes} are independent of each other in all temperature and magnetic field ranges. This means, in the considered configuration there is no coupling between ferromagnetic and exchange modes. This can be explained by the symmetry considerations. The system represented by a ferrimagnet with uniaxial anisotropy and magnetic field applied along anisotropy axis has axial symmetry. Eigenmodes of axially-symmetric system remain circular both far and close to $T_M$, which means no hybridization occur:

\begin{equation}
    \begin{aligned}
        \varphi_{\substack{\mathrm{fm}\\ \mathrm{ex}}} &=A_j e^{-\tfrac{t}{\tau_j}} \sin \left(\omega_jt+\psi_j\right)\\
        \theta_{\substack{\mathrm{fm}\\ \mathrm{ex}}} &=\pm A_j e^{-\tfrac{t}{\tau_j}} \cos\left(\omega_jt+\psi_j\right), 
    \end{aligned}
    \label{Eq: trajectories}
\end{equation}
where $j$ denotes corresponding "fm" or "ex" mode, and the sign describes opposite directions  of $\mathbf{L}$ vector rotation in ferromagnetic and exchange modes. Here and in the following equations in signs $\pm$ the upper sign corresponds to the fm-mode and the bottom sign - to the ex-mode. Eqs.~\eqref{Eq: trajectories} correspond to $m>0$ case, while for $m<0$ the rotation direction of both modes will be opposite with respect to $\mathbf{L}$ vector due to the opposite alignment of $\mathbf{L}$ and $\mathbf{M}$ vectors below and above $T_M$.

\subsection{Frequencies of the eigenmodes}

Analytical solution of Eqs. (2-3) for complex frequencies of these modes, i.e. taking into account magnetic damping, is:
\begin{align}
    \omega_\mathrm{fm}^\mathrm{complex}&=\sqrt{\Omega^2+\left(\delta - i\Gamma \right)^2}+\left(\delta - i\Gamma \right),\label{Eq: complex_omega_fm}\\
    \omega_\mathrm{ex}^\mathrm{complex}&=\sqrt{\Omega^2+\left(-\delta - i\Gamma \right)^2}+\left(-\delta - i\Gamma \right).
    \label{Eq: complex_omega_ex}
\end{align}

If the magnetic damping is relatively small, i.e. ${\Gamma\ll\omega_0}$, then real part of the frequencies can be written as:
\begin{align}
    \omega_{\substack{\mathrm{fm}\\ \mathrm{ex}}} &=\omega_0 \pm \delta \label{Eq: omega},
    \end{align}
as shown in Fig.~\ref{Fig: freqs}a where both spin-mode frequencies are shown at the frequency diagram with respect to the temperature $T$ and magnetic field $H$. Note that at the compensation point ${\delta = \gamma H\geq0}$.

\begin{figure*}[htb]
(a)~~~~~~~~~~~~~~~~~~~~~~~~~~~~~~~~~~~~~~~~~~~~~~(b)~~~~~~~~~~~~~~~~~~~~~~~~~~~~~~~~~~~~~~~~~~~~~
(c)~~~~~~~~~~~~~~~~~~~~~~~~~~~~~~~~~~~~~~~~~~~~~~~~~~~~~~~~~
\\
  \includegraphics[height=0.3\linewidth]{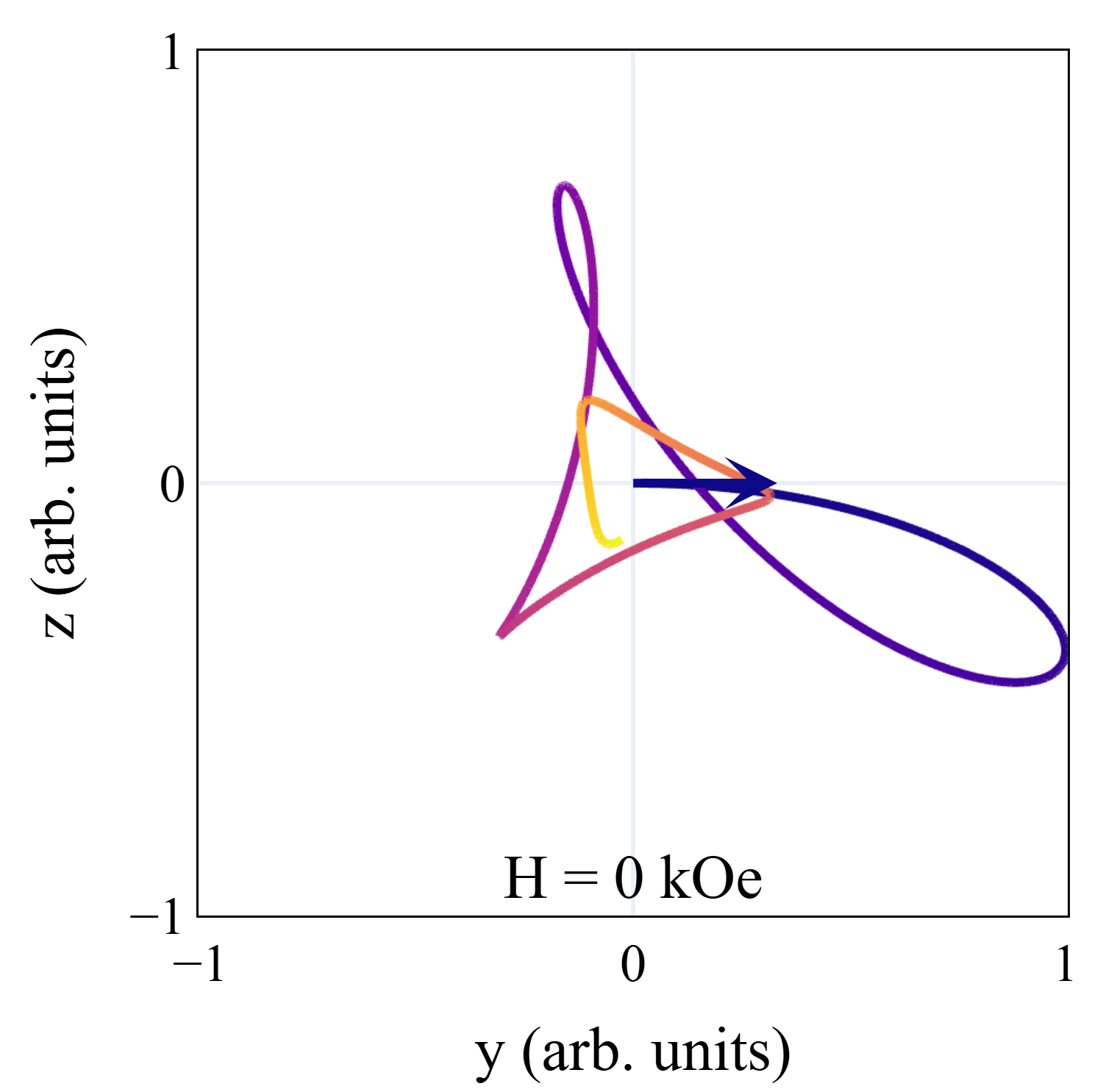}\hfill
  \includegraphics[height=0.3\linewidth]{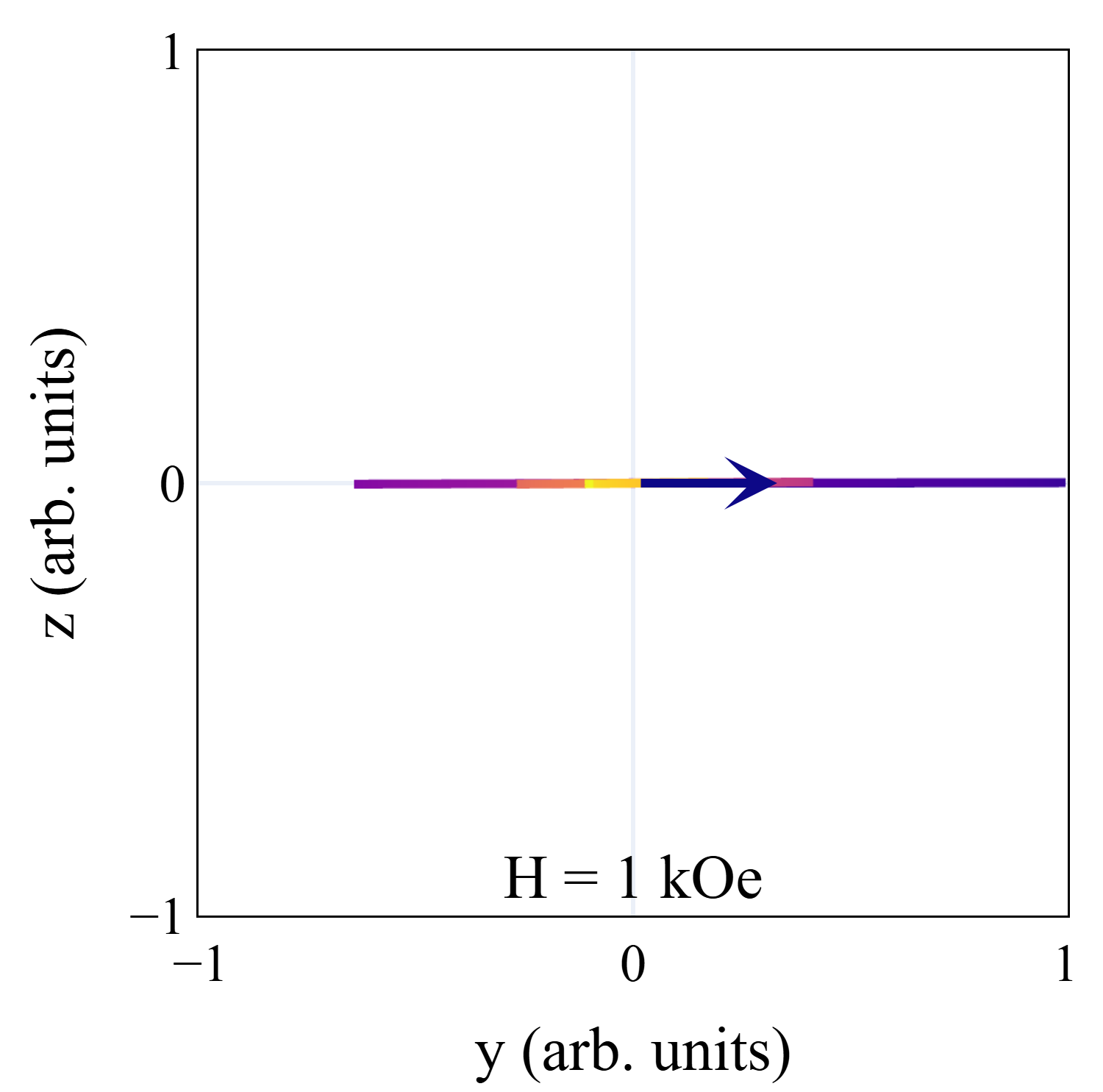}\hfill
  \includegraphics[height=0.3\linewidth]{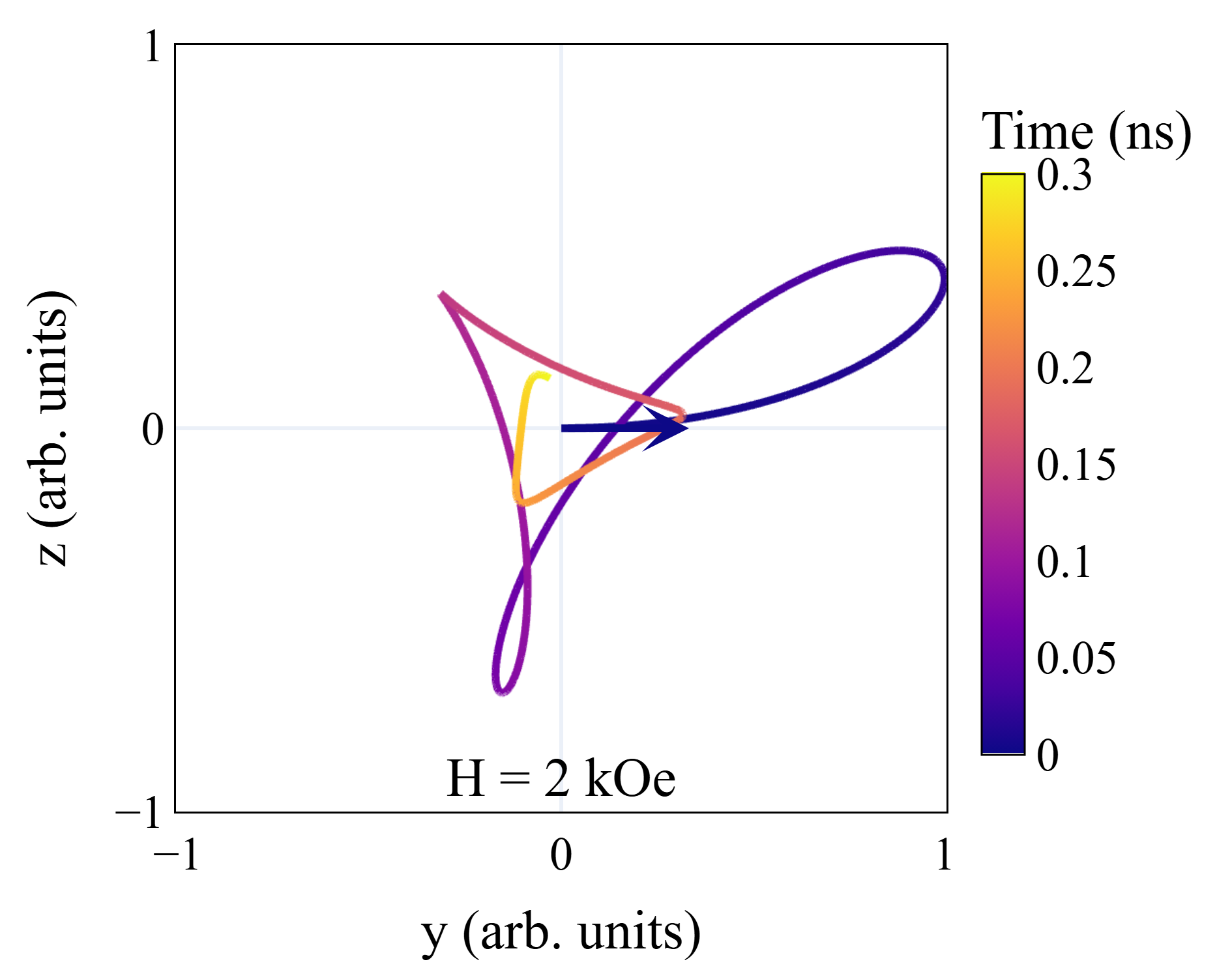}

  \caption{Neel vector $\mathbf{L}$ trajectories at $T=325.1$~K temperature and (a) {$H=0$~kOe} (ferromagnetic mode has lower frequency) (b) {$H=1$~kOe} (frequencies of exchange and ferromagnetic modes are equal) (c) {$H=2$~kOe} (exchange mode has lower frequency) (see Fig.~\ref{Fig: freqs}a). The arrow shows the arrival direction of the pump pulse in the YZ-plane.}
  \label{Fig: L_traj}
\end{figure*}

According to Eq.~\eqref{Eq: omega}, far from the compensation point the frequency of the exchange mode is much higher than of the ferromagnetic mode. At the same time, Fig.~\ref{Fig: freqs}a shows that in the vicinity of the compensation point mode frequencies become closer to each other. Moreover, the frequency surfaces shown in $(T,H)$ coordinates cross each other at the condition $\delta= 0$, or:
\begin{equation}
    H=\frac{\left|m\right|}{2\chi_\perp},
    \label{Eq: eqv_condition}
\end{equation}
which means that the frequencies of exchange and ferromagnetic modes become equal at this condition $\omega_\mathrm{ex}=\omega_\mathrm{fm}=\omega_0$. Eq.~\eqref{Eq: omega} provides linear dependence of $\omega_j$ on $H$:

\begin{equation}
     \omega_{\substack{\mathrm{fm}\\ \mathrm{ex}}}=(\omega_0 \mp \frac{1}{2} \omega_\mathrm{KK}) \pm \gamma H,\label{Eq: omega_lin} 
\end{equation}
which means that both modes are linearly dependent on the external magnetic field with equal-value and opposite-sign derivatives. This is quite an unusual situation compared to the configuration when the external magnetic field is applied perpendicular to the magnetic anisotropy axis: for the temperatures far from $T_M$ the exchange mode weakly depends on the external magnetic field, and for the temperatures  close to $T_M$ both frequencies demonstrate weak and nonlinear dependence on the magnetic field axis~\cite{krichevsky2023unconventional}. Moreover, Eq.\eqref{Eq: omega_lin} predicts that $\omega_\mathrm{ex}\rightarrow0$ under the condition that the magnetic field is strong enough.

\section{Excitation of spin dynamics by fs-laser pulses}\label{sec:excitation}

\subsection{Experimental}

To excite spin dynamics in the ferrimagnetic film experimentally we used circularly polarized femtosecond laser pulses. Due to the inverse Faraday effect each laser pulse influence on spins almost instantaneously, at the scale of the pulse duration by a kind of torque which deviates the spins from their equilibrium state. Microscopically this process is attributed to the impulsive stimulated Raman scattering of photons on magnons. It can be described phenomenologically by an effective magnetic field $h_\mathrm{IFE}$ which influences on spins during the laser pulse propagation through the magnetic sample: $\mathbf{h}_\mathrm{IFE} \propto \left[\mathbf{E}\times\mathbf{E}^\ast\right]$.
Therefore, $\mathbf{h}_\mathrm{IFE}$ is directed along the wavevector of the laser pulse and its value maximizes for the circular polarization.
 
For the experimental studies we use the pump-probe method, with the optical pump and probe beams incident at quite a large angle of 65° on the iron-garnet film. Large angles of incidence were chosen due to the out-of-plane configuration of the applied magnetic field.

Spin dynamics was excited via the inverse Faraday effect (IFE) induced by the circularly-polarized 180-fs femtosecond pump pulses of 800-nm wavelength (see Appendix A for the details). Measurement of spin dynamics was performed using the transient Faraday rotation of a linearly polarized femtosecond probe pulse of 525-nm wavelength delayed with respect to the pump pulse. The pump and probe pulses were focused on the sample to the spots with $13~\mathrm{\mu m}$ and $9~\mathrm{\mu m}$ (fluence of $0.5~\mathrm{mJ/cm^2}$ and $0.1~\mathrm{mJ/cm^2}$) diameters correspondingly. The obtained transient Faraday rotation signals were approximated by two decaying harmonic functions (see Appendix B for the details). One should note that exchange and ferromagnetic notation of experimental points is provided with respect to the theoretical predictions, since only the projection of the mode precession on the probe k-vector is measured, so that the rotation direction couldn't be determined in this scheme.

\begin{figure*}[htb]
~~~~(a)~~~~~~~~~~~~~~~~~~~~~~~~~~~~~~~~~~~~~~~~~~~~~(b)~~~~~~~~~~~~~~~~~~~~~~~~~~~~~~~~~~~~~~~~~~~~~(c)~~~~~~~~~~~~~~~~~~~~~~~~~~~~~~~~~~~~~~~~~~~~~(d)~~~~~~~~~~~~~~~~~~~~~~~~~~~~~~~~~~~~~~~~~~~~~~~~~~~~~~\\
  \includegraphics[height=0.23\linewidth]{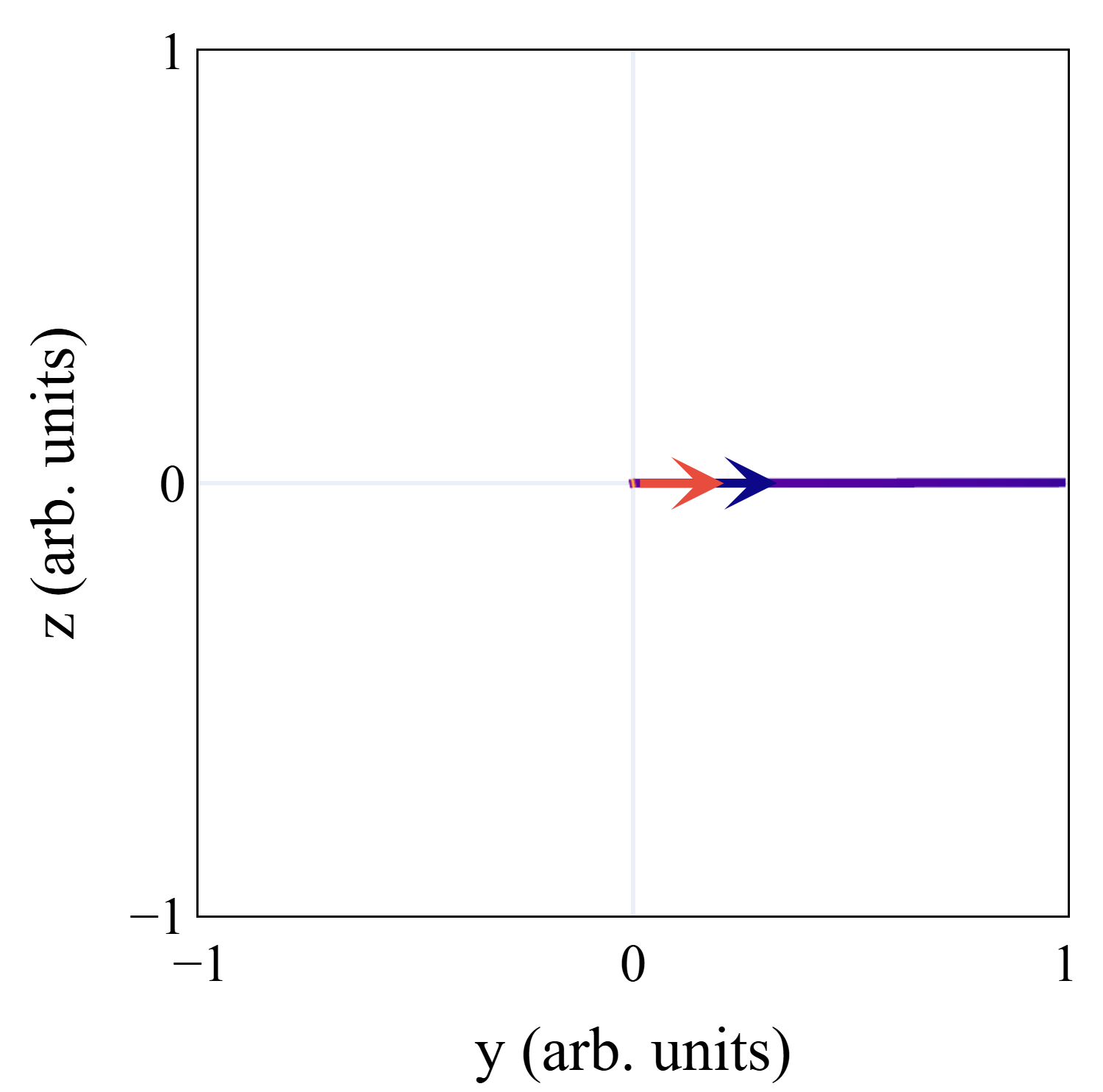}\hfill
  \includegraphics[height=0.23\linewidth]{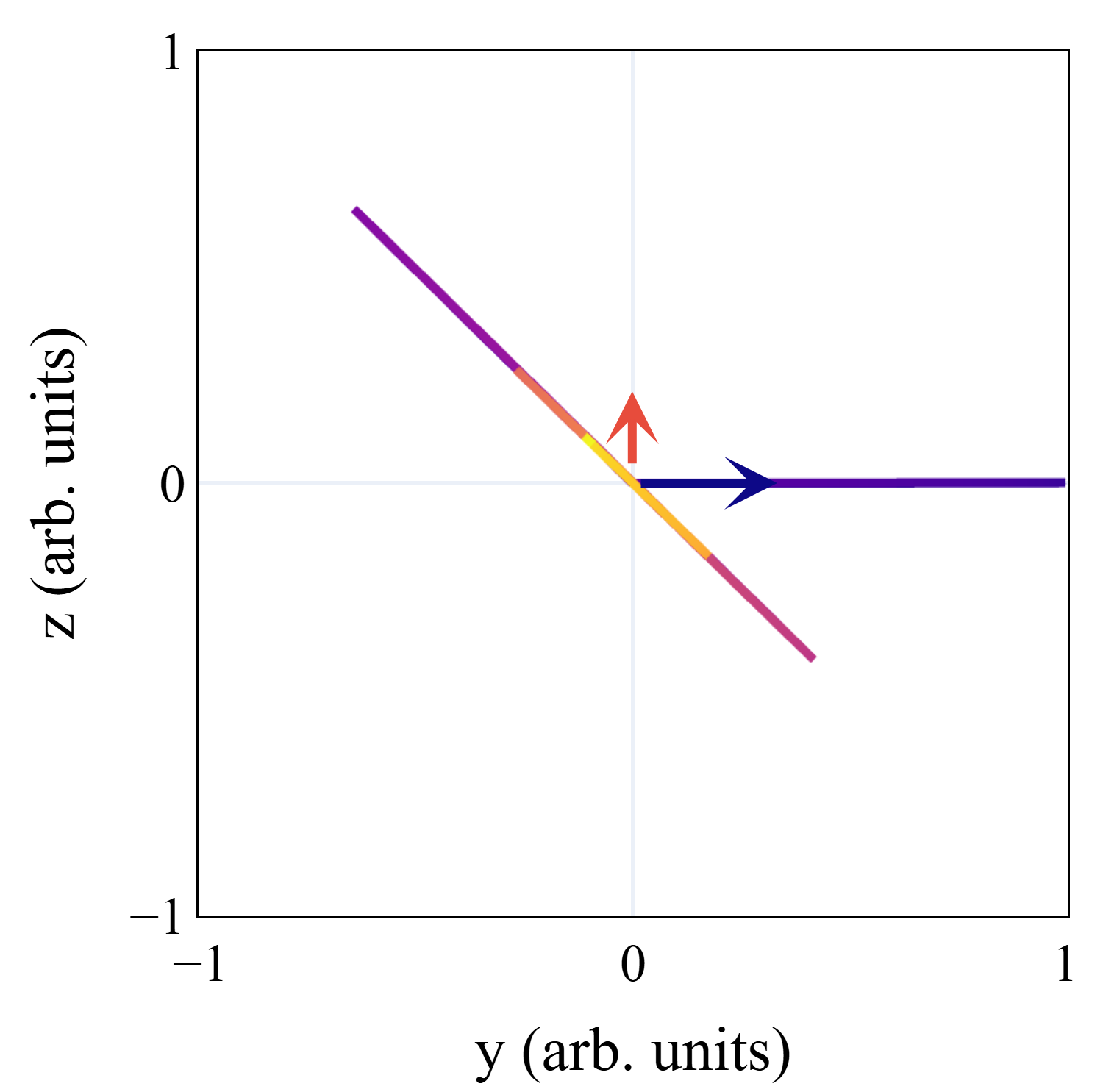}\hfill
  \includegraphics[height=0.23\linewidth]{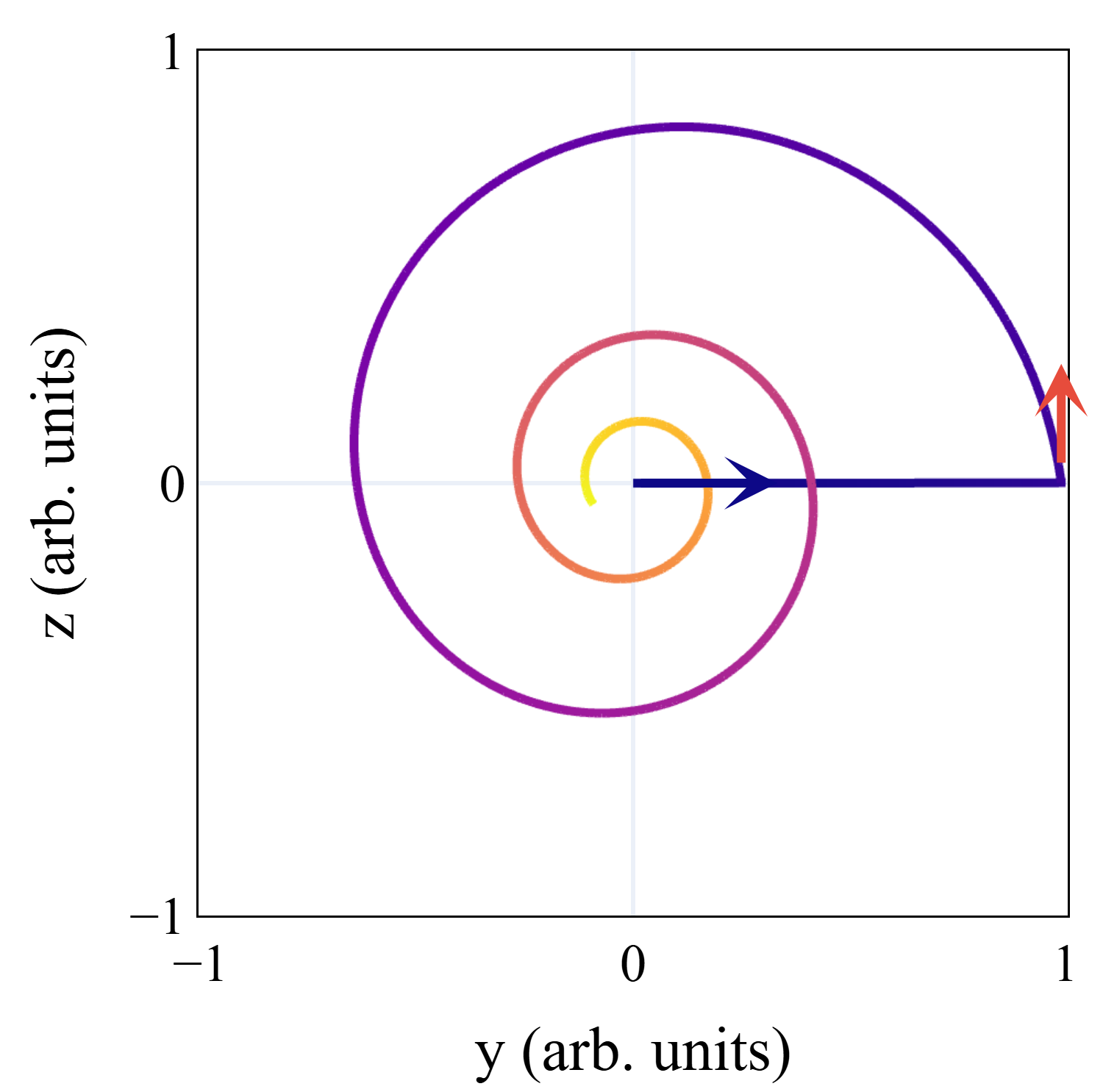}\hfill
  \includegraphics[height=0.23\linewidth]{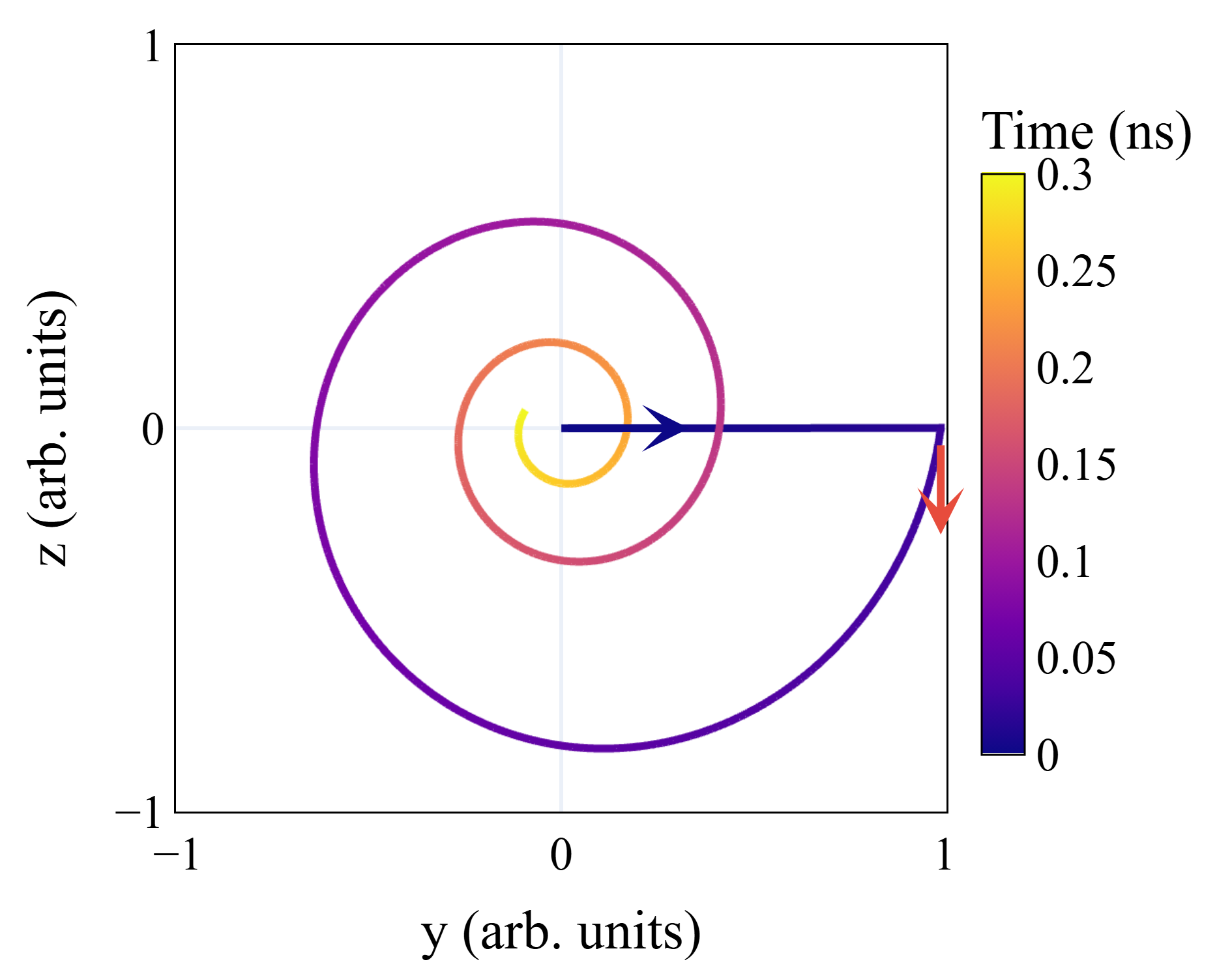}

  \caption{Two-pulse control of the Neel vector $\mathbf{L}$ trajectory in the case of equal mode frequencies (Fig.~\ref{Fig: L_traj}(b)) at $T=325.1$~K and $H=1$~kOe. The blue arrow shows the arrival direction of the first pulse. The red arrow indicates the arrival direction of the second pulse, its position along the trajectory marks the arrival time. The length of arrows is proportional to the value of the effective field of the inverse Faraday effect $h_\mathrm{IFE}$ and marks the magnitude of the knocks. (a) A half-period-delayed knock is applied against the motion. (b) A half-period-delayed knock is applied perpendicular to the oscillation axis. (c, d) A second knock is applied after a quarter period $T/4$ perpendicular to the oscillation plane. Trajectory color encodes time.}
  \label{Fig: two_pulses}
\end{figure*}

Experimental measurements shown in Figs.~\ref{Fig: freqs}b-d are in a good agreement with the numerical calculations and confirm the theoretical predictions discussed above. Fig.~\ref{Fig: freqs}b shows that $f_\mathrm{ex}$ and $f_\mathrm{fm}$ become equal at {$T\approx318$~K, which is 15~K below the magnetization compensation point $T_\mathrm{M}$. At the room temperature $T=298$~K the mode frequencies differ greatly ($f_\mathrm{ex}\sim35$~GHz and $f_\mathrm{fm}\sim8$~GHz at small $H$, see Fig.~\ref{Fig: freqs}c). With an increasing of the magnetic field $H$ they only tend to each other linearly, as the theory predicts the intersection at $H=5.8$~kOe which is well above the magnetic fields available in the setup. At the temperature of $T=320$~K (Fig.~\ref{Fig: freqs}d), the intersection of frequency curves at $H\approx1$~kOe is vividly seen. Theory predicts the intersection point at somewhat larger field of $H\approx1.8$~kOe (see inset to (Fig.~\ref{Fig: freqs}d) which is due to the not quite precise values of the magnetic parameters. Nevertheless, linear dependencies of both mode frequencies on $H$ with opposite-sign derivatives nicely correspond to experimental data.

\subsection{Trajectories of modes} 

Another important point is the trajectory of the magnetization and Neel vector motion under different conditions. As the action of the femtosecond pulse is short in terms of characteristic times of the spin motion, it can be described as a photonic knock~\cite{ignatyeva2025high}. Due to the IFE-induced magnetic field is directed along the wavevector of pump laser pulse lying in the XY-plane, launching of spin dynamics might be simulated as the initial conditions in the form $\theta|_{t=0}=\varphi|_{t=0}=\dot\theta|_{t=0}=0$ and $\dot\varphi|_{t=0}=\gamma(\omega_H+\omega_\mathrm{KK})h_\mathrm{IFE}\tau$, where $h_\mathrm{IFE}$ is the value of the IFE-induced magnetic field, and $\tau$ is the duration of the femtosecond pulse  (Fig.~\ref{Fig: config}a).

The $\mathbf{L}$-vector trajectories were numerically simulated using Eqs.~\eqref{Eq: theta_and_phi} with the initial conditions described above. It was found out that the initial phases of both ferromagnetic and exchange modes (Eq.~\eqref{Eq: trajectories}) are equal zero: ${\psi_\mathrm{ex}=\psi_\mathrm{fm}=0^\circ}$, and their magnitudes and quality factors are equal to each other as well: ${A_\mathrm{ex}=A_\mathrm{fm}},~{Q_\mathrm{ex}=Q_\mathrm{fm}}$. These relationships do not depend on the choice of magnetic field and temperature.

Figure~\ref{Fig: L_traj} shows the trajectories for 3 different external magnetic field values. The frequency ratio of ferromagnetic and exchange modes indeed changes, which might be seen in Figure~\ref{Fig: L_traj}a,c as swapping the rotation direction of high-frequency and low-frequency modes. Besides, at the point where the frequencies become equal, such features lead to the degeneration of the resulting complex precession trajectory to an oscillation along the effective magnetic field of IFE, which coincides with the k-vector of the exciting pulse.

This is quite an exotic shape for the spin motion in ferrimagnets. The resulting trajectories of $\mathbf{L}$-vector calculated at the fixed temperature of 325.1 K at different magnetic fields below, equal and above the magnetic field at which the frequencies of the two spin modes coincide is shown in Fig.~\ref{Fig: L_traj}.

\subsection{Two-pulse precession control} 

To demonstrate the opportunity to control the trajectory of precession launched by an ultrashort optical excitation, we consider a two-pulse protocol applied to the case shown in Fig.~\ref{Fig: L_traj}(b), where two frequencies coincide. Fig.~\ref{Fig: two_pulses} presents the effect of a second photonic knock with different magnitudes and directions. The arrow position marks the delay time of the second pulse. Furthermore, the arrow indicates the direction and relative magnitude of the second knock with respect to the first.

Action of the second pulse after half-period of the Neel vector precession $T/2$ corresponds to the moment when the Neel vector passes the equilibrium position with the velocity opposite to the one provided by the first IFE pulse. Thus, if the second pulse is equivalent to the first and arrives at this moment, it gives a velocity opposite to the one that the Neel vector has. This results in the immediate stopping of the spin motion (Fig.~\ref{Fig: two_pulses}a). Note that if precession damping is high enough, the second pulse should be $e^{-T/2\tau}$ times smaller in magnitude.

Perpendicular direction of the second pulse arriving at the same $T/2$ interval adds the orthogonal component to L velocity. It changes the orientation of the axis along which oscillations occur, so it turns at $\tan^{-1}\left(e^{T/2\tau}h_\mathrm{IFE, 2}/h_\mathrm{IFE, 1}\right)$ angle. By tuning the magnitude of the second pulse, one might control the oscillation axis and turn it in arbitrary direction (Fig.~\ref{Fig: two_pulses}b)). Moreover, third, forth, etc. pulses acting after integer number $n$ of half-periods $nT/2$, might turn the oscillation axis further.

More interesting is situation when the second pulse acts after quarter-period $T/4$ and has perpendicular to the first direction (see Fig.~\ref{Fig: two_pulses}c, d). In this case, it acts when the initial planar oscillation is at the turning point and the velocity created by the first pulse is zero. The second pulse then excites the orthogonal oscillation direction, while the delay between the pulses make the motion trajectory circular. According to Eq.~\eqref{Eq: trajectories}, depending on the direction of the second pulse, its action is in-phase with one of ferromagnetic or exchange mode, and antiphase with the other. Thus, second pulse action results in the supression of one of the spin modes and circular precession corresponding to pure other mode. Note that helicity of the laser pulse switches IFE direction, and provides selectivity for either ferromagnetic or exchange mode launching.

\section{Conclusion}\label{sec:conc}

We investigated spin dynamics in a nearly compensated uniaxial ferrimagnet in the presence of a magnetic field applied along the magnetic anisotropy axis. Experimental measurements and numerical modeling demonstrate that, near the magnetization compensation point, the frequencies of the two ferrimagnetic eigenmodes approach each other and become strongly sensitive to the external magnetic field. The modes correspond to clockwise and counterclockwise rotations of the Néel vector and exhibit equal-magnitude but opposite-sign field dependences, resulting in a critical condition where their frequencies coincide and the rotation handedness of both modes reverses.

Analysis of the spin trajectories reveals an unusual dynamical regime occurring at mode degeneracy. The characteristic two-frequency precessional motion transforms into a linear oscillation whose direction is determined by the inverse-Faraday-effect excitation. Furthermore, we show that excitation by a second femtosecond pulse enables manipulation of the spin trajectory, providing control over the resulting motion without modification of the sample geometry.

These findings demonstrate that nearly compensated ferrimagnets support unconventional dynamical states that emerge from the interplay of exchange interaction, magnetic anisotropy, and external magnetic field near compensation. The revealed effects provide additional opportunities for controlling spin dynamics and magnon motion in ferrimagnetic systems and may be relevant for optically driven magnonic applications.

\begin{acknowledgments}
This research was funded by the Russian Science Foundation, project No.23-62-10024.
\end{acknowledgments}

\bibliography{my_paper}

\newpage
\clearpage

\section*{Appendix A. Pump-induced inverse Faraday effect}

Oblique incidence of a circularly-polarized fs pulse disturbs its polarization so it becomes elliptical. However, even at the angle of 65° according to the Fresnel formulas the ratio of the ellipse semi-axes is 1:0.73, so that polarization exciting spin dynamics can be considered close to circular. A state of optical polarization insight the magnetic film influences the values of the instantaneous torque by light exerted on on spins. The latter can be calculated by an effective magnetic field $\mathbf{h}_\mathrm{IFE}$ describing the inverse Faraday effect: $\mathbf{h}_\mathrm{IFE} \propto \left[\mathbf{E}\times\mathbf{E}^\ast\right]$.

Experimental pump-probe scheme is based on the modulation of the pump polarization using a photoelastic modulator and further measurement of the dynamics at the corresponding modulation frequency. This excludes influence of any of the polarization-independent or linear-polarization-related effects. 

In addition, we have measured the precession signals for the two opposite orientations of the external magnetic field $\mathbf{H}$, see Fig.~\ref{Fig: IFE_proof}. Identical dynamics pattern proves the absence of $H$-odd effects (such as inverse Coutton-Moutton effect).

\begin{figure}[htb]
  \includegraphics[width=\columnwidth]{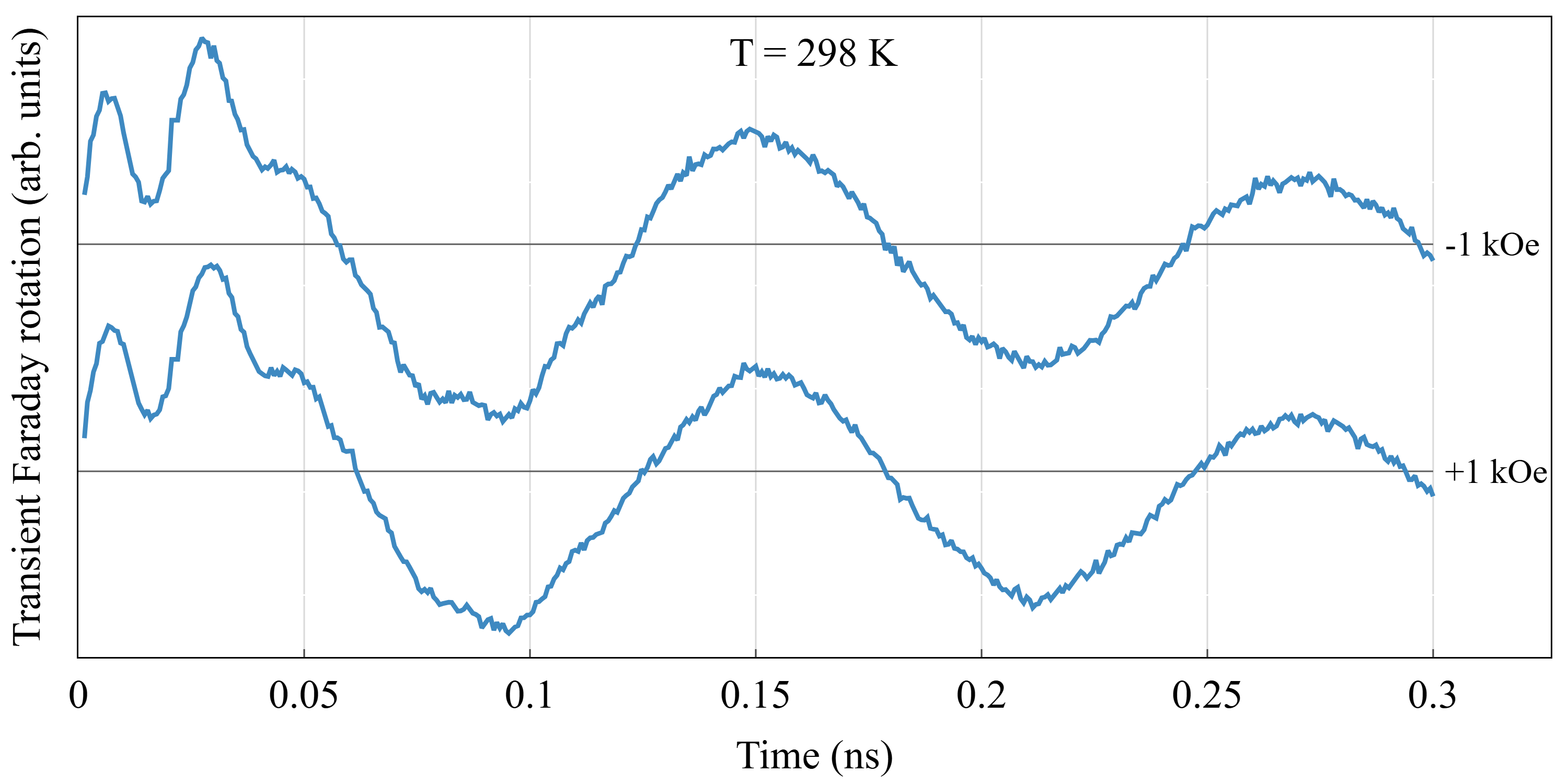}
  \caption{Spin dynamics measured for $+\mathbf{H}$ and $-\mathbf{H}$ directions.}
  \label{Fig: IFE_proof}
\end{figure}

\section*{Appendix B. Approximation of transient Faraday rotation signals}

The obtained transient Faraday rotation signals were approximated by two decaying harmonic functions:
\begin{align*}
    \Phi(t)=&A_1 e^{-t/\tau_1} \sin(2\pi f_1 t +\psi_1) + \\
    &A_2 e^{-t/\tau_2} \sin(2\pi f_2 t +\psi_2)
\end{align*}

As a first step, an FFT-based spectral estimate was used to identify dominant frequency regions and to generate initial parameter guesses for nonlinear fitting. However, when the eigenfrequencies are well separated, the exchange mode often decays much faster than the ferromagnetic mode; therefore, its spectral contribution may be strongly suppressed or even buried in the noise floor in the global FFT spectrum.

To recover such rapidly decaying components, a continuous wavelet transform (CWT)~\cite{torrence1998cwt} was additionally employed, since it provides time-localized spectral information and enables inspection of short-time intervals where the exchange mode is still observable.

For the case of closely spaced modes, where spectral overlap limits resolution, the ESPRIT~\cite{roy1989esprit} subspace method was used to obtain high-resolution frequency estimates.

Although FFT/CWT and ESPRIT are emphasized for different spectral regimes, in practice these methods were applied jointly and in parallel to produce an extended set of initial parameter guesses. These candidates were then passed to nonlinear least-squares regression (Levenberg–Marquardt algorithm) for the full damped-harmonic model, and the final solution was selected as the best-quality approximation according to the fitting residual and parameter consistency. Signals and their approximations are shown in Fig.~\ref{Fig: Signal approximation}.

In all the cases, root mean square error (RMSE) was calculated to verify that bi-harmonic functions provided statistically better approximations than single-frequency one (Fig.~\ref{Fig: 1vs2_modes}). 

\begin{figure}[htb]
  \includegraphics[width=\columnwidth]{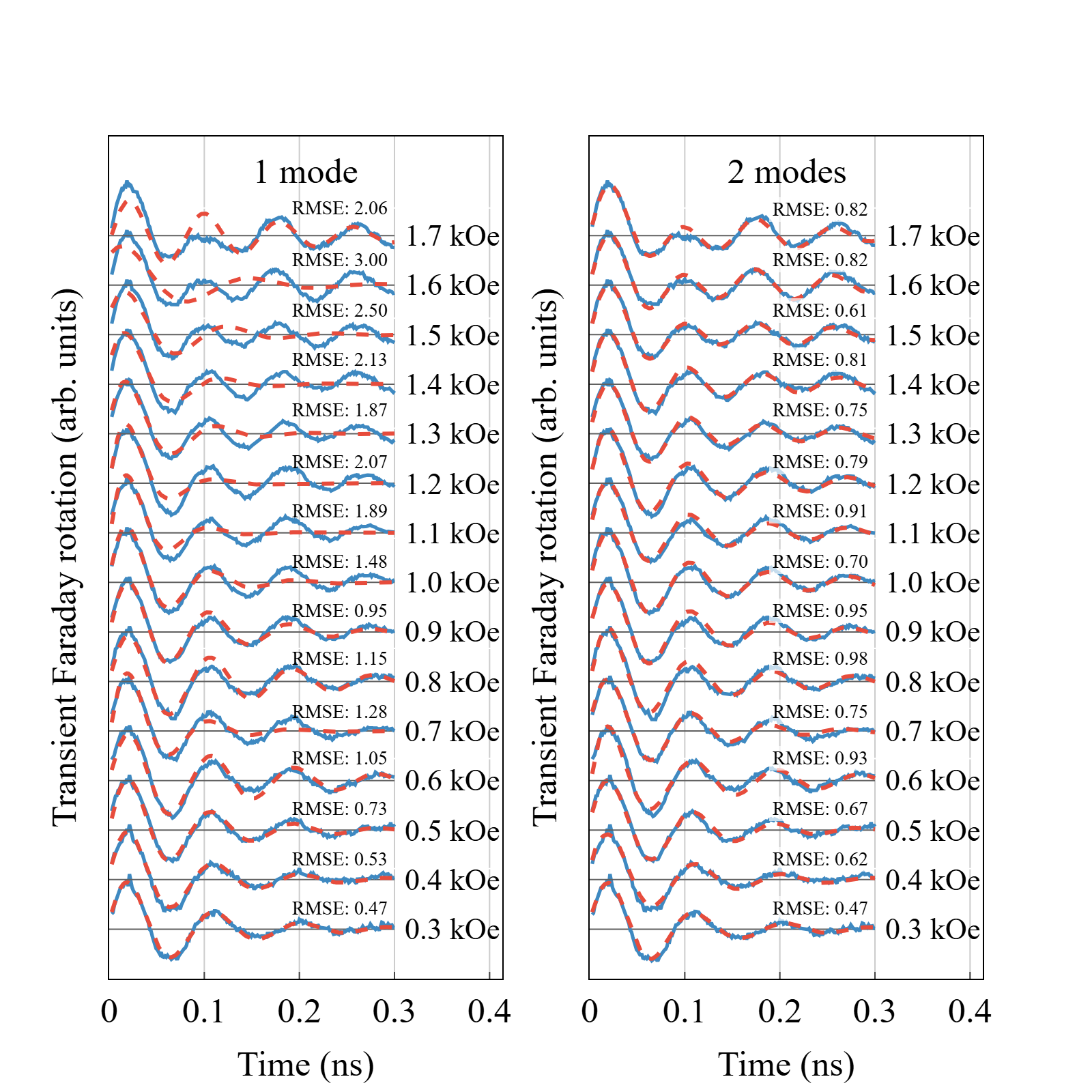}
  \caption{Normalized TFR signals at $T=320$~K and their approximations with one or two modes. ${RMSE=\sqrt{\tfrac{\sum_{points}(Signal-Fit)^2}{N_{points}}}}$.}
  \label{Fig: 1vs2_modes}
\end{figure}

\begin{figure*}[htb]
(a)~~~~~~~~~~~~~~~~~~~~~~~~~~~~~~~~~~~~~~~~~~~~~~~~~~~~~~~~~(b)~~~~~~~~~~~~~~~~~~~~~~~~~~~~~~~~~~~~~~~~~~~~~~~~~~~~~~~~~
(c)~~~~~~~~~~~~~~~~~~~~~~~~~~~~~~~~~~~~~~~~~~~~~~~~~~~~~~~~~
\\
  \includegraphics[width=0.32\linewidth]{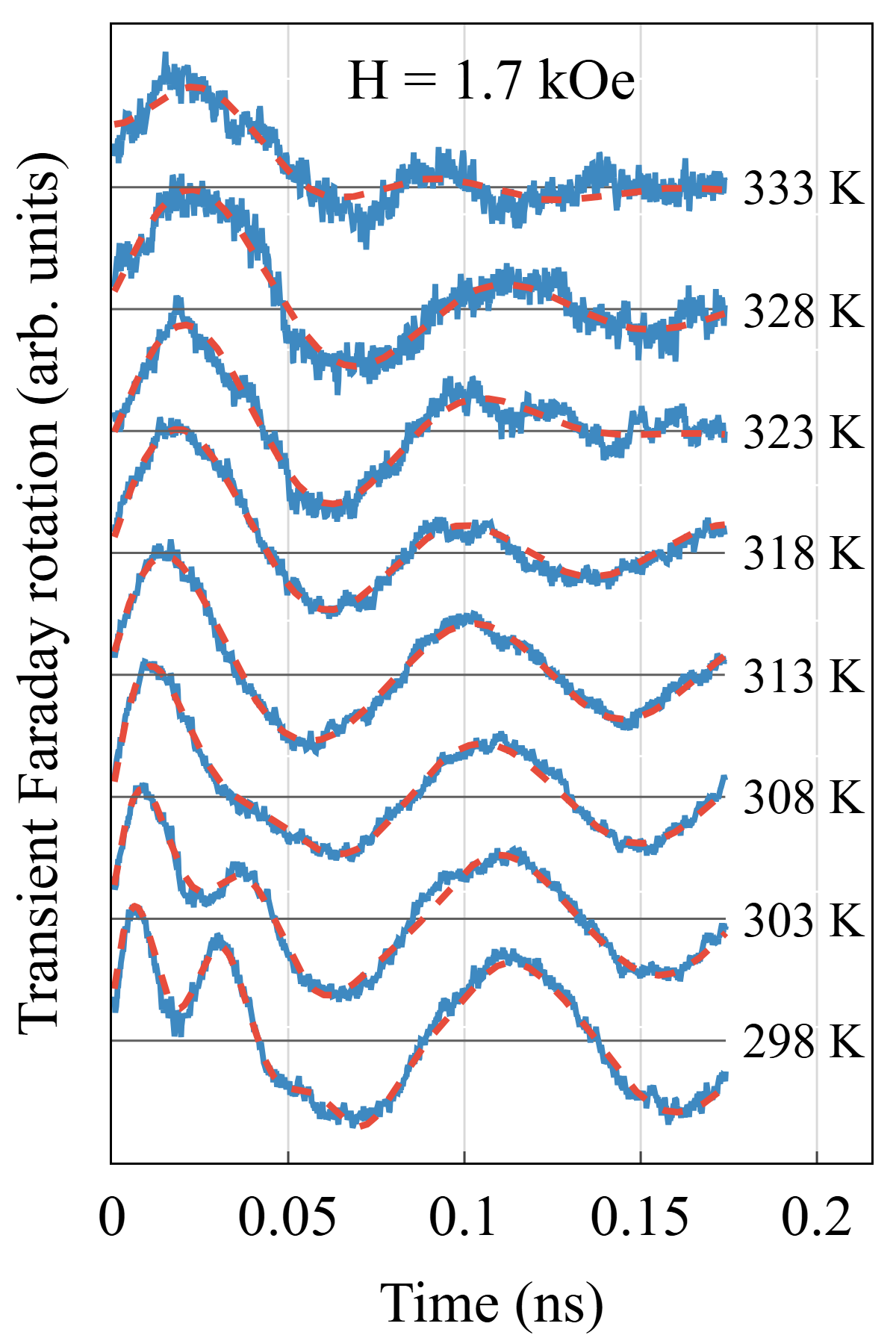}
  \includegraphics[width=0.32\linewidth]{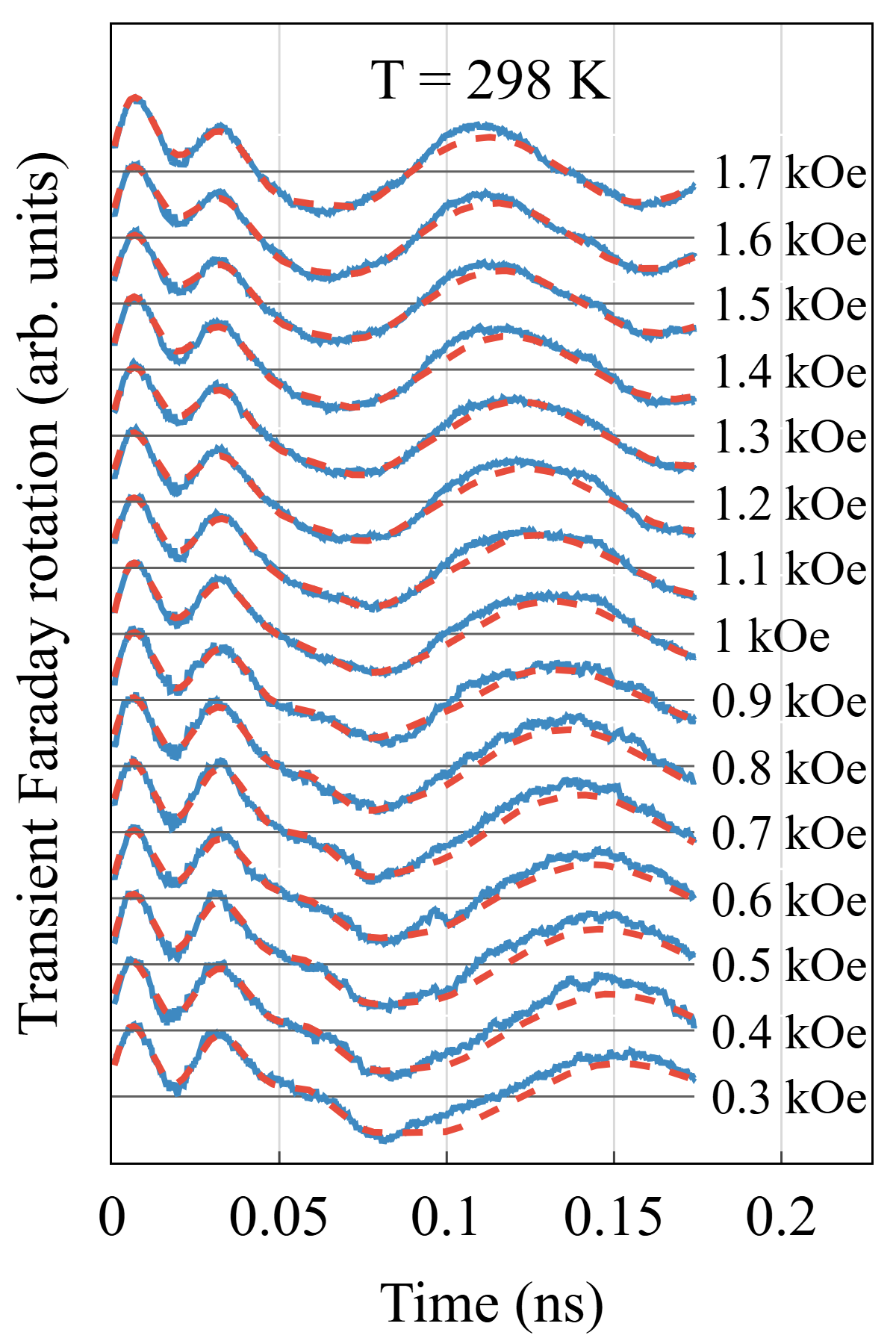}
  \includegraphics[width=0.32\linewidth]{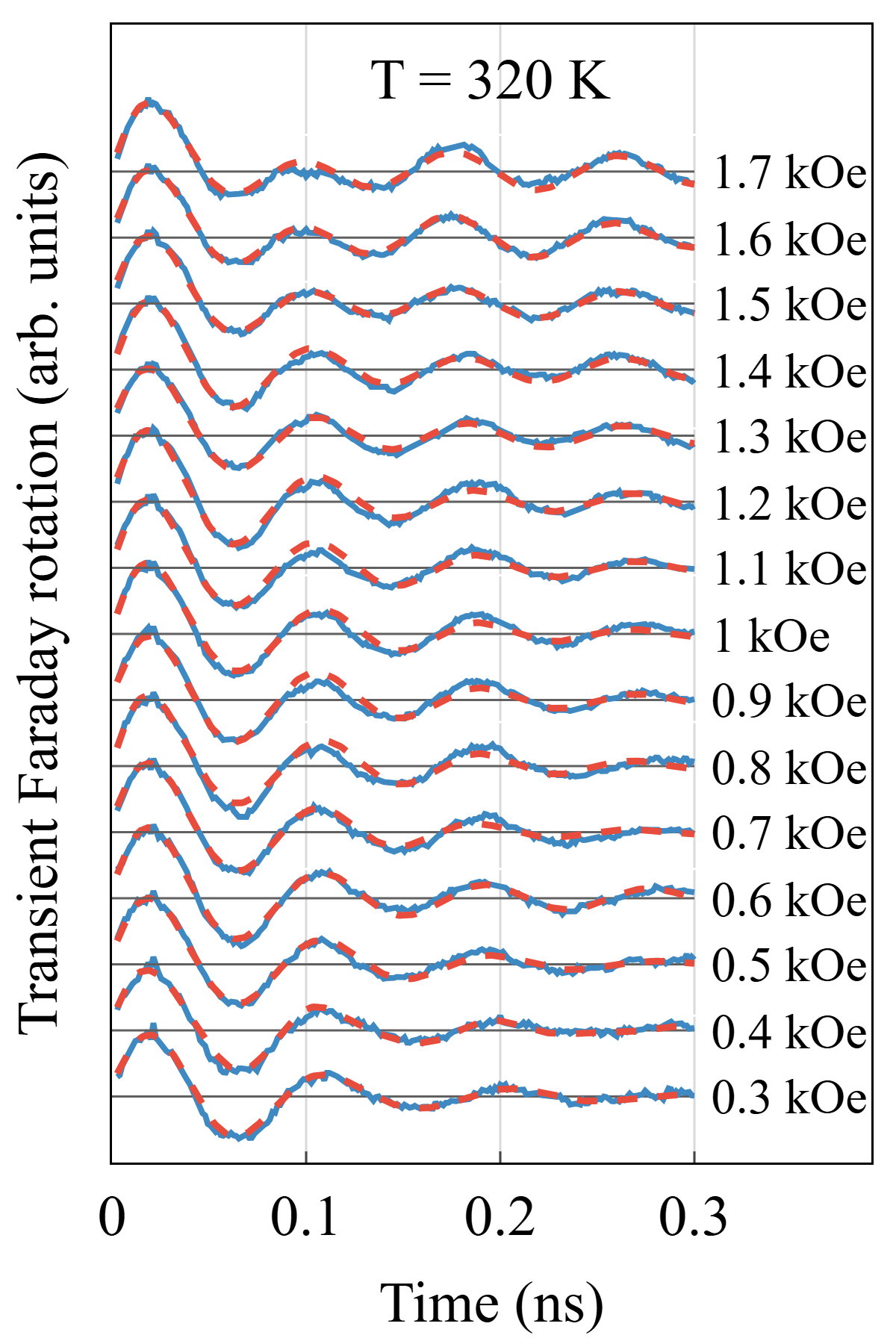}
    
  \caption{Normalized TFR signals and their approximations for the cases shown in Fig.~\ref{Fig: freqs}b-d: (a) $H=1.7$~kOe; (b) $T=298$~K; (c) $T=320$~K;}
  \label{Fig: Signal approximation}
\end{figure*}

\end{document}